\documentclass[10pt]{emulateapj}
\usepackage{bm}
\usepackage{pslatex}
\usepackage{graphicx}
\usepackage{natbib}
\begin{document}
\title{Whistler mode waves and the electron heat flux in the solar wind: Cluster observations}

\author{C.~Lacombe\altaffilmark{1}, O.~Alexandrova\altaffilmark{1}, L.~Matteini\altaffilmark{2}, O.~Santol\'ik\altaffilmark{3,4}, N.~Cornilleau-Wehrlin\altaffilmark{1,5}, A.~Mangeney\altaffilmark{1}, Y.~de~Conchy\altaffilmark{1}
 and M.~Maksimovic\altaffilmark{1}}

\altaffiltext{1}{LESIA, Observatoire de Paris, PSL Research University, CNRS, UPMC Universit\'e Paris 06, Universit\'e Paris-Diderot, 5 place Jules~Janssen, F-92190 Meudon, France.}
\altaffiltext{2}{Imperial College, London, SW7 2AZ, UK.}
\altaffiltext{3}{Institute of Atmospheric Physics ASCR, Prague, Czech Republic.}
\altaffiltext{4}{Faculty of Mathematics and Physics, Charles University in Prague, Czech Republic.}
\altaffiltext{5}{LPP, CNRS, Ecole Polytechnique, UPMC, Route de Saclay, F-91128 
Palaiseau, France.}

\begin{abstract}
The nature of the magnetic field fluctuations in the solar wind between the ion and electron scales is still under debate. Using the Cluster/STAFF instrument, we make a survey of the power spectral density and of the polarization of these fluctuations at frequencies $f\in[1,400]$~Hz, during five years (2001--2005), when Cluster was in the free solar wind. In $\sim 10\%$ of the selected data, we observe narrow-band, right-handed, circularly polarized fluctuations, with wave vectors quasi-parallel to the mean magnetic field, superimposed on the spectrum of the permanent background turbulence. We interpret these coherent fluctuations as whistler mode waves. The life time of these waves varies between a few seconds and several hours. Here we present, for the first time, an analysis of long-lived whistler waves, i.e. lasting more than five minutes.  We find several necessary (but not sufficient) conditions for the observation of whistler waves, mainly a low level of the background turbulence, a slow wind, a relatively large electron heat flux and a low electron collision frequency.  When the electron parallel beta factor $\beta_{e\parallel}$ is larger than 3, the whistler waves are seen along the heat flux threshold of the whistler heat flux instability. The presence of such whistler waves confirms that the whistler heat flux instability contributes to the regulation of the solar wind heat flux, at least for $\beta_{e\parallel} \ge$ 3,  in the slow wind, at 1 AU.  
\end{abstract}


\maketitle

\vspace{0.5cm}

\section{Introduction}\label{sec:intro}

 The electromagnetic fluctuations in the frequency range $\sim[1,500]$~Hz have been studied in the solar wind since decades, see e.g. \citet{beinroth81,denskat83,lengyel-frey94,lengyel-frey96,lin98}, and the review by \citet{briand09}. The nature of these fluctuations, however, is still under debate. The mentioned frequency range is sometimes called  {\it whistler range} \citep{beinroth81,denskat83} because it corresponds to fluctuations below the electron  cyclotron frequency $f_{ce}$,  where the whistler wave mode may exist. 

 \citet{lengyel-frey96} and \citet{lin98} have observed whistler modes in a large range of heliographic latitudes and distances from the Sun,  using the Ulysses/URAP measurements of the electric and magnetic field spectra. Whistler modes have been  as well found near interplanetary shocks ({\it e.g.} \citet{pierre95,wilson09}) and at the Earth's bow shock \citep{hoppe81,elaoufir90}.

However, the {\it whistler frequency range}  is populated by the permanent magnetic field turbulence. This latter  has a negligible frequency in the solar wind frame,  is  Doppler shifted  in the spacecraft frame, and therefore is observed in the same frequency range as whistlers.  

As the whistler mode waves are polarized coherent waves,  polarization measurements are thus needed to separate   the waves from the background turbulence.
The lack of polarization observations can lead to erroneous interpretations. For instance, some properties of supposed whistler waves, like a power law frequency spectrum or a correlation between the wave intensity and the magnetic  field strength \citep{beinroth81,lengyel-frey96}, are probably mainly properties of  the background solar wind turbulence, to which whistlers can, or cannot, be superimposed. Conversely, the possible presence of whistlers demands particular care when investigating the permanent solar wind turbulence. For instance, some of the spectral breaks or knees shown by \citet{sahraoui13a,sahraoui13b} should not be considered as characteristic features of the permanent turbulence, as long as they are due to the superimposition of a narrowband of intermittent whistler waves.

The observation of coherent electric field and/or magnetic field waveforms, and polarization measurements, are thus necessary to confirm the whistler wave mode identification. Such observations have been made on Geotail \citep{zhang98}, on WIND \citep{moullard01} and on STEREO \citep{breneman10}. In these papers, the polarization of the waves was deduced from the waveforms of a Time Domain Sampler (TDS) or of a Wave Form Capture instrument. The limitations of these measurements is their threshold in amplitude and their short time recording: the waveforms measured by Geotail last 8 seconds; those on Wind last from 20~ms up to 0.1~s. These measurements reveal bursts of narrowband and short-lived whistler modes, in propagation quasi-parallel to the mean magnetic field ${\bf B_0}$,  in the free solar wind, in the electron foreshock of the Earth's bow shock \citep{zhang98}, and in magnetic clouds \citep{moullard01}.
 Using the electric field STEREO/TDS data, \citet{breneman10}  have made an automatic survey, during two years (2007-2009), of 10 minutes groups of the most intense polarized events, lasting 0.15~s, with at least one event par minute. The authors found that these groups of intense oblique whistler waves appear mostly within the stream interaction regions and close to interplanetary (IP) shocks. 

In this paper we provide the first continuous observations of long-lived whistler waves in the free solar wind using the Cluster satellites. The four Cluster spacecraft cruise around the Earth, from the solar wind to the magnetosphere, with an apogee  $\simeq$~20~$R_E$ and a perigee $\leq$~4~$R_E$.  The STAFF experiment on Cluster gives the power spectral density of the magnetic field fluctuations from about 1 Hz  to $f_{ce}\simeq 300$~Hz, and above. It gives also, continuously, the polarization of these fluctuations (every 4~s).  We analyze five years of Cluster data (2001-2005) and we select  time intervals of free solar wind (i.e., not magnetically connected to the Earth's bow-shock). Then, within these intervals we separate the polarized fluctuations from the non-polarized ones. 

The non-polarized fluctuations ($\sim 90\%$ of the selected data) have been studied by \citet{alexandrova09a,alexandrova12}. These fluctuations have a general spectral shape between the ion scales  and a fraction of electron scales. The intensity of these spectra is well correlated to the ion thermal pressure $nkT_{p}$ \citep{alexandrova13a}. These non-polarized electromagnetic fluctuations seem to have a negligible frequency in the solar wind frame, and a wave-vector anisotropy $k_{\perp}\gg k_{\|}$ (paper in preparation). In the spacecraft frame, they are Doppler shifted in the {\it whistler range}. 

The present study is focused  on  the rest $\sim 10\%$ of the selected data, which show a clear right-handed (RH) polarization with respect to ${\bf B_0}$, and a propagation direction of the fluctuations quasi-parallel to the magnetic field. We interpret these fluctuations as quasi-parallel whistler mode waves. The lifetime of these waves lasts from seconds up to several hours. We look for the solar wind properties which favor the presence of  long-lived whistlers,  i.e. coherent waves observed during more than 5 minutes. We also consider the electron heat flux and the electron temperature anisotropy for these intervals. Note that whistlers are not a permanent feature: in a region where they are observed, they can be intermittent.

\section{Instruments and data}

The present study relies on data sets from different experiments onboard the Cluster fleet. The Spatio-Temporal Analysis of Field Fluctuations (STAFF) experiment on Cluster  \citep{cornilleau-wehrlin97,cornilleau-wehrlin03} measures the three orthogonal components of the magnetic field fluctuations in the frequency range $0.1$~Hz--$4$~kHz, and comprises two on board analysers, a wave form unit (STAFF-SC) and a Spectrum Analyser (STAFF-SA). STAFF-SC provides the digitised wave form up to either 10 or 180 Hz, depending on the spacecraft telemetry rate. The Spectrum Analyser uses the three magnetic field components and two electric field components (from the EFW experiment, \citet{gustafsson97}) to build a 5x5 spectral matrix every 4~s, between 8~Hz and 4~kHz (in the normal telemetry rate). Then, the PRopagation Analysis of STAFF-SA Data with COherency tests (the PRASSADCO program) gives the wave propagation properties every 4~s by a Singular Value Decomposition (SVD) of the spectral matrix \citep{santolik03}. Both experiments, STAFF-SC and STAFF-SA, allow to determine the polarization sense, the ellipticity and the propagation  direction of the fluctuations observed in the frequency range of the whistler mode waves: indeed,  the maximum of the electron gyrofrequency $f_{ce}$ is  of the order of 500~Hz in the solar wind at 1~AU, below the upper limit of the STAFF-SA frequency range.  The use of the electric field components gives the sense of the wave vector ${\bf k}$, without the $180^{\circ}$ ambiguity of the direction of the normal to the polarization plane of the magnetic fluctuations \citep{santolik01,santolik03}.  However, the electric field data are not always good between 8 and 30~Hz:  this is due to artefacts in the wake of the spacecraft, in the solar wind \citep{eriksson06,lin03}. Thus, the sense of ${\bf k}$ is not always clear below  30~Hz.

The WHISPER experiment \citep{decreau97}  is used to check that Cluster is in the free solar wind,  {\it i.e.} that the magnetic field line through Cluster does not intersects the Earth's bow shock: there is no electrostatic or Langmuir wave, typical  of the foreshock. Some of the used data are available at the CSDS (Cluster Science Data System): the magnetic field ${\bf B_0}$, given every 4~s by the FGM experiment \citep{balogh97}; 
the proton density $N_p$, the wind  velocity $V_{sw}$  and the proton temperature $T_p$ parallel and perpendicular to ${\bf B_0}$ derived from the CIS/HIA  experiment data \citep{reme97}. The electron parameters  given by the Low Energy Electron Analyser of the PEACE experiment \citep{johnstone97}  are taken from the CAA (Cluster Active Archive): in the following we  use the electron   temperatures $T_{e\parallel}$ and $T_{e\perp}$, parallel and perpendicular to ${\bf B_0}$, and the heat flux vector,  ${\bf Q}_e$.  The electron temperatures are the total electron temperatures; the heat flux is the total electron heat flux.  A separation between the core, halo and strahl populations, see {\it e.g.} \citet{stverak08,stverak09},  should be done in a next step. As for the electron density $N_e$, we shall assume that it is equal to $N_p$. 

For the STAFF, FGM and CIS experiments, we mainly consider the Cluster~1 data. For the electron parameters, we 
use the Cluster spacecraft with the highest resolution data, generally Cluster~2 or Cluster~ 4. 
 The fact that the data come from different spacecraft is
not a drawback. Indeed, we only look at intervals with relatively small spacecraft separations,  i.e. from 2001 to 2005,  so that the STAFF wave data are very similar on the four spacecraft.

\section{Data selection}

We have explored the Cluster data from 2001 to 2005, when the separation between the spacecraft was smaller than 3000 km. We have considered six months every year, from December to May, when Cluster is able to sample the free solar wind.  The fact that Cluster is in the free solar wind, not magnetically connected to the Earth's bow shock, is deduced from the absence of the electrostatic waves typical of  the electron foreshock. It is confirmed by the calculation of the depth of the  spacecraft in the foreshock, for a paraboloid model of the Earth's bow shock \citep{filbert-kellogg79}, as was done for example by \citet{lacombe85} and \citet{alexandrova13b}.

As explained by \citet{alexandrova12}, the orbit of Cluster implies that the angle 
$\Theta_{BV}$ between the ${\bf B_0}$ field and  the solar wind velocity ${\bf V_{sw}}$ is larger than  $60^{\circ}$ in intervals of free solar wind. 

We have selected intervals of 10 minutes, giving spectra of magnetic field fluctuations averaged over 10 minutes. 
When Cluster was continuously in the free solar wind 
for several hours, we have only selected about one interval every hour. In this way, 
we obtain 175 independent intervals, on 30 different days.
Among these 175 intervals, 149 display the usual magnetic field turbulence of the solar wind, 
made of non-polarized fluctuations with a smooth spectrum, without bumps or knees; their spectral 
shape has been analyzed by \citet{alexandrova12}. These fluctuations have a negligible frequency 
in the solar wind frame.  Their wave vectors ${\bf k}$ are mainly perpendicular to the average magnetic field ${\bf B_0}$, with a quasi-gyrotropic distribution: this can be shown (paper in preparation) by an analysis similar to the one of \citet{bieber96}, \citet{mangeney06}  and \citet{alexandrova08c}.

The 26 other intervals display polarized fluctuations and spectral bumps at frequencies where the polarized fluctuations are observed. These fluctuations can either last during the considered 10~minutes, or can be made of intermittent bursts lasting less than a few minutes. With the condition that the polarized fluctuations last more than 5 minutes, we obtain a sample of  20 intervals where the polarized fluctuations can be considered as well established.

 Our data set is not very large, but represents well the free solar wind at 1~AU. Indeed we have tried to select time intervals with different plasma conditions. We could have built a larger data set by considering not only one interval of 10 minutes every hour but all the intervals in the free solar wind. However, several consecutive intervals, which have nearly identical properties, would not have really enriched our set of the solar wind properties.


\begin{figure}
\includegraphics[width=9cm]{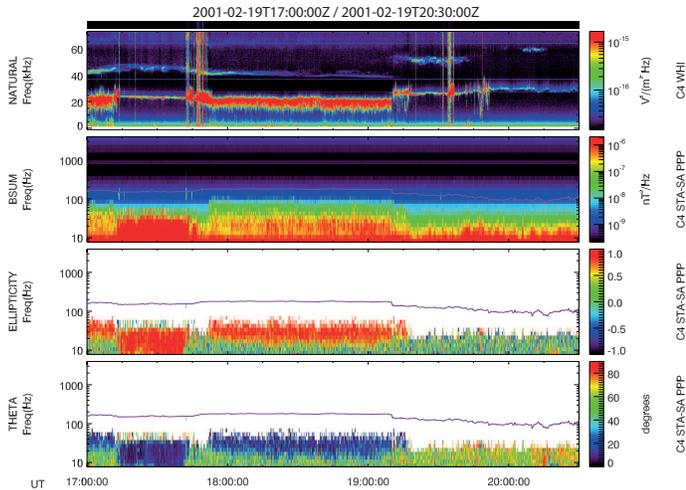}
\caption{Day 2001/02/19, Cluster-4, spectra and polarization, from the foreshock to the free solar wind (see text). Panel 1: electric field dynamic spectrum (WHISPER) from $4$ to $80$~kHz. Panel 2: dynamic spectrum of the total energy of magnetic fluctuations from 8~Hz to 4~kHz. Panel 3: ellipticity of the magnetic fluctuations. Panel 4: the angle $\Theta_{kB}$ between the direction ${\bf k}$ perpendicular to the polarization plane  and the field ${\bf B_0}$. The continuous line in panels 2--4 gives the value of the electron gyrofrequency.}
\label{fig1}
\end{figure}

The dynamic spectra of Figure~\ref{fig1} illustrate the intensity and the polarization of  fluctuations found in the free solar wind (and in the foreshock) on 2001/02/19. The upper panel gives the intensity of the electric  field fluctuations observed 
by WHISPER from 4 to 80 kHz: the intense fluctuations around the electron plasma frequency $f_{pe} \simeq$ 25-30 Hz indicate that Cluster is in the foreshock  during the time interval[17:00--17:15]~UT,  then during  [17:42--19:20]~UT, and intermittently from 19:20 to  
19:50~UT. Cluster is in the free solar wind  during the interval [17:15--17:42]~UT. Here, the intense magnetic fluctuations observed by STAFF-SA below about 40~Hz 
(panel 2) are whistler mode waves: indeed, their ellipticity close to 1 (panel 3) indicates a quasi-circular right-handed polarization. (For a left-handed polarization,  the ellipticity is $-1$;  a linear polarization corresponds to an ellipticity close to zero \citep{santolik01,santolik03}). The polar angle $\Theta_{kB}\simeq 0^{\circ}$ between the wave vector ${\bf k}$ and ${\bf B_0}$ (panel 4)  implies a quasi-parallel propagation.  The polarized fluctuations in the foreshock, before 17:15~UT and from 17:42 to 19:20~UT are  whistler waves  as well (see section 10). For the interval [19:50--20:15]~UT, Cluster is again in the free solar wind  without signatures of  polarized waves (panel 3): the magnetic fluctuations are the usual solar wind non-polarized turbulence. 

The free solar wind interval [17:30--17:40]~UT belongs to the 20 intervals with well-established whistlers in our sample; the interval  [20:00--20:10]~UT belongs to the 149 intervals of usual non-polarized turbulence, studied by \citet{alexandrova12,alexandrova13a}. The properties of the polarized fluctuations and the conditions of their  occurrence are analyzed in the following sections,  where, in addition to the selected 10 minutes intervals, we shall consider three longer time intervals, when whistlers appear and last half an hour, or more,  and two short time intervals with whistlers observed by STAFF-SC in  a high telemetry mode (up to 180~Hz) hereafter called burst mode.

\section{Frequency and wave number of the observed waves}
\begin{figure}
\includegraphics[width=9.0cm]{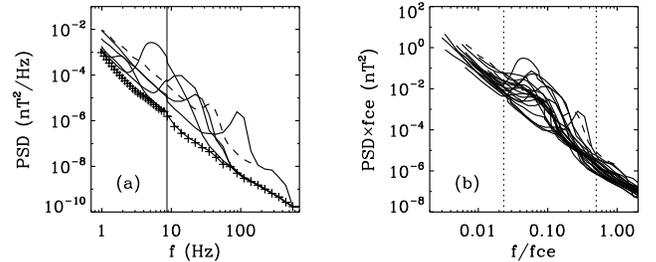}
\caption{Spectra of solar wind magnetic field fluctuations with whistler bumps, measured on Cluster~1.  (a) Six selected spectra (power spectral density versus frequency). The crosses give the background noise on STAFF  measured in the magnetospheric lobe, on 2004/08/12, [11:00--11:10]~UT; the vertical solid line at 8.5~Hz corresponds to the  junction between the SC and SA units on the STAFF instrument (in normal mode);  dashed line: average over 1 minute of whistlers observed by STAFF-SC in burst mode.  (b)~Solid lines: 20 spectra, averages over 10 minutes, with long-lived whistlers, versus $f / f_{ce}$;  dashed line: same as in (a); the vertical dotted lines correspond to $f_{lh}$ and $0.5f_{ce}$ (see text).}
\label{fig2}
\end{figure}


In Figure~\ref{fig2}(a), six selected spectra on Cluster~1 are shown, giving the Power Spectral Density (PSD),  the sum of the intensities of the magnetic field fluctuations in the three directions. We see that spectral bumps,  found to correspond to the polarized fluctuations, can be observed from 1 to 200~Hz.  The crosses give the background noise. The five solid line spectra are normal mode spectra  averaged over 10 minutes, using both STAFF-SC and STAFF-SA.  Below 8.5~Hz (vertical solid line) the spectra are Morlet wavelet spectra \citep{torrence98}  of the STAFF-SC waveforms. Above 8.5 Hz, the spectra are averages of the STAFF-SA spectra (the discontinuity at 8.5~Hz indicates that the inter-calibration between SC and SA is good but not perfect, probably because a poor calibration at 8.8, 11 and 14~Hz on Cluster~1 gives a  slightly underestimated signal on STAFF-SA).  The dashed line spectrum is a wavelet spectrum of the STAFF-SC waveforms in burst mode,  averaged over 1 minute (day 2009/01/31, 04:52-04:53~UT). 

In Figure~\ref{fig2}(b), the 21 spectra (20 in normal mode and 1 in burst mode) with polarized fluctuations are drawn as functions of the ratio $f / f_{ce}$. We see that the bumps are observed between   the lower hybrid frequency $f_{lh} \approx (f_{ce} f_{ci})^{1/2}$  and $0.5f_{ce}$
(vertical dotted lines),  where $f_{ci}$ is the proton gyrofrequency: this frequency range is typical of whistler mode waves. The refractive index n of the whistler mode in cold plasmas can be approximated as (Baumjohann and Treumann, 1996, eq. 9.155): 
\begin{equation}
 n^2 =  \frac{k^2 c^2}{\omega^2} \approx 1 + \frac {\omega_{pe}^2}{\omega (\omega_{ce} cos\Theta_{kB} - \omega)},
\end{equation} 
\noindent where $\Theta_{kB}$ is the angle between the wave vector and the field ${\bf B_0}$, and $\omega_{pe}$ the electron plasma frequency. In the right hand term of this equation, 1 is negligible in the solar wind. The wave number $k$ can then easily be estimated with the observed frequency  as
 \begin{equation}\label{eq:k-w}
 \frac {k^2 c^2}{\omega_{pe}^2}  \approx  
 \frac {\omega^2}{\omega (\omega_{ce} cos\Theta_{kB} - \omega)}.
\end{equation} 
\noindent Assuming that $\Theta_{kB}$ is very small (as we shall see in the next section), eq.~(\ref{eq:k-w})  gives the wave numbers corresponding to the frequencies of Figure~\ref{fig2}(b): $kc/ \omega_{pe}$ varies between 0.1 and 0.9, and  $k r_{ge}$ varies between 0.1 and 0.8, where $r_{ge} = \sqrt{2 k_B T_{e\perp}}/ \omega_{ce}$ is the electron gyroradius.

\section{Polarization and direction of the wave vectors}

\begin{figure}
\includegraphics[width=6.cm]{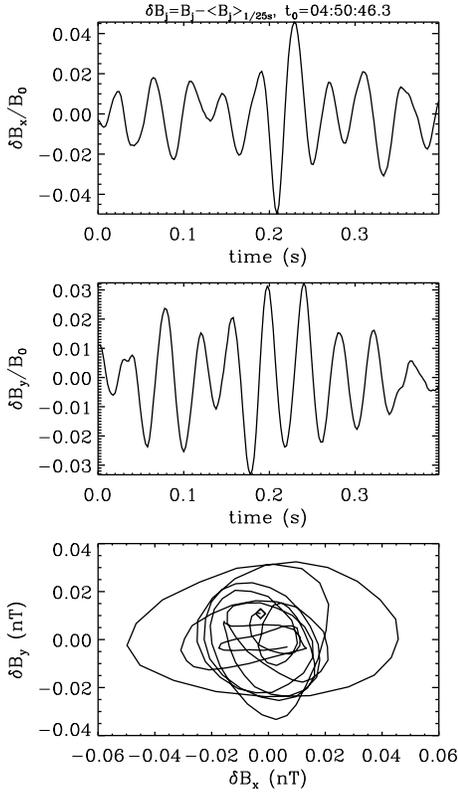}
\caption{Coherent whistler waves observed by Cluster-2/STAFF-SC on 31 of January 2009 at 04:50:46~UT during the time interval of the spectrum in Figure~1c of \citet{sahraoui13b}. Two upper panels: perpendicular normalized magnetic field fluctuations at 25~Hz, defined as $\delta B_j=B_j-\langle B_j \rangle_{\tau=0.04\;\;s}$ in the local field aligned frame. Lower panel: polarization in the plane perpendicular  to ${\bf B_0}$; the beginning of the hodogram is indicated by a diamond.} 
\label{fig:sc-waves}
\end{figure}

 
Figure~\ref{fig:sc-waves} shows an example of a coherent whistler waveform lasting less than 1~s, measured by Cluster-2/STAFF-SC  in burst mode and corresponding to the time interval of a spectrum with a break around 25~Hz, published in Figure~3 of \citep{sahraoui13a} and in Figure~1(c) of \citep{sahraoui13b}.  The two upper panels give magnetic fluctuations at 25~Hz in the plane perpendicular to the mean magnetic field. The bottom panel gives the polarization in this plane: it is quasi-circular and right handed with respect to ${\bf B_0}$, which is aligned with ${\bf z}$ here.  
This event (not belonging to our sample of long-lived whistlers) represents an example of intermittent whistlers in a narrow frequency band superimposed on the background turbulence spectrum. In this particular case, these whistler waves produce a spectral break around 25~Hz.

\begin{figure}
\includegraphics[width=8cm]{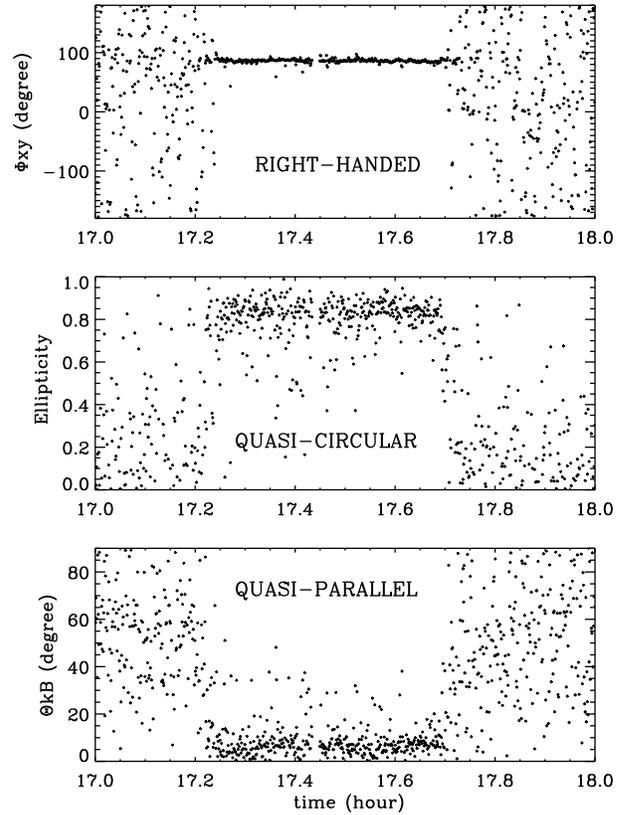}
\caption{Day 2001/02/19 on Cluster~4. The SVD analysis of the spectral matrix of the magnetic fluctuations at 14 Hz, every 4 s, gives 
from top to bottom: the phase angle $\Phi_{xy}$, the ellipticity (related to the ratio of the two largest eigenvalues), 
and the angle $\Theta_{kB}$ between the wave vector ${\bf k}$ and the field ${\bf B_0}$.}
\label{fig:pol}
\end{figure}


Figure~\ref{fig:pol} displays long-lived whistlers lasting about half an hour, observed by  Cluster-4/STAFF-SA  in normal mode (at the beginning of the interval shown in Figure~\ref{fig1}). 
 Here we show three characteristic parameters of the waves observed in a spectral bump, around $0.1f_{ce}$ (14~Hz). 
The upper panel gives the phase difference $\Phi_{xy}$ between the fluctuations measured at 14~Hz in two orthogonal directions $x$ and $y$, perpendicular to ${\bf B_0}$. $\Phi_{xy} \simeq  90^{\circ}$ implies a right-handed polarization,  observed  during [17:15--17:42]~UT. (Rotations of ${\bf B_0}$ occurred at 17:15 and 17:42~UT, so that Cluster~4 leaves the Earth's foreshock to enter  the free solar wind at 17:15~UT, and re-enters the foreshock at 17:42~UT, see Figure~\ref{fig1}). The second panel gives the ellipticity  (see \citet{maksimovic01,santolik01,santolik03} for the exact definition), related to the ratio of the axes of the polarization ellipse: an ellipticity larger than 0.8 implies a quasi-circular polarization. The third panel gives the angle $\Theta_{kB}$ between the wave vector ${\bf k}$ and ${\bf B_0}$. This angle is less than $10^{\circ}$, implying a quasi-parallel propagation. This observation of waves with small but nonzero values of $\Theta_{kB}$ are in fact consistent with a gyrotropic distribution of wave vectors having a maximum probability density in the direction of the local magnetic field line ($\Theta_{kB} \approx 0^{\circ}$), taking into account the fact that the probability density  would be equal to $\sin\Theta_{kB}$ if the distribution was isotropic.

We conclude that the whistler waves observed on this day have a right-handed quasi-circular polarization in the spacecraft frame, with a direction of propagation quasi-parallel to ${\bf B_0}$. 
 All whistler intervals of our sample display the same wave properties, right-handed and quasi-circular polarization, with a  quasi-parallel propagation.

\section{Visibility of the whistlers}

\begin{figure}
\includegraphics[width=9.0cm]{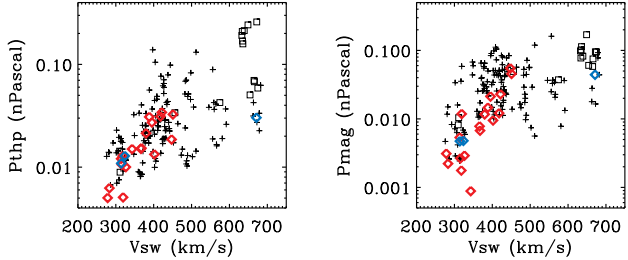}
\caption{
Left panel: scatter plot of the proton thermal pressure $P_{thp}$ versus the solar wind  speed $V_{sw}$ for 170 intervals. The red (blue) diamonds are the 18 (3) intervals with intense (weak) whistler waves. The black symbols indicate the 149 intervals without whistlers (usual background turbulence), with squares for the intervals downstream of interplanetary shocks, crosses for the other intervals. Right panel: the solar wind magnetic pressure $P_{mag}$ versus  $V_{sw}$, for the same sample.}
\label{fig:visib}
\end{figure}


While the magnetic field turbulence with a  regular  spectrum  is a permanent feature of the solar wind, the whistler waves are not permanent. Let us look for the solar wind conditions when whistler waves are observed.   In the left panel of Figure~\ref{fig:visib}, the 149 intervals without whistlers are represented by black symbols  (crosses or squares) in the plane ($V_{sw},P_{thp}$) where $V_{sw}$ is the solar wind speed, and $P_{thp} = N_p k_B T_p$ the  mean proton thermal pressure in nPascal, over 10 minutes. The black squares represent intervals downstream  of interplanetary (IP) shocks. The red diamonds represent the 18 intervals with intense enough  whistlers,  i.e., the waves with an energy 4 times higher than the usual solar wind turbulence measured on the same day at the same frequency. The blue diamonds represent the 3 intervals where whistlers are less intense. We note that intense whistler waves can be observed when $V_{sw}$ is less than 500~km/s, and when $P_{thp}$ is below 0.04~nPa.

What can be the reasons for these visibility conditions of the whistlers?

A first reason is that, when $V_{sw}$ is large, the spectrum of the usual solar wind turbulence suffers a large Doppler shift.  Indeed,  as far as the turbulent wave vectors ${\bf k}_t$ are mainly perpendicular to ${\bf B_0}$, with a gyrotropic distribution,  and $\Theta_{BV}$ is large (see section 3), some wave vectors ${\bf k}_t$ make a small angle with $V_{sw}$, yielding a large Doppler shift. Conversely, the whistler wave vectors make a small angle with ${\bf B_0}$, and thus a large angle with $V_{sw}$, yielding a small Doppler shift. The consequence is that, when $V_{sw}$ is large, the frequency shift of the  regular turbulence spectrum can mask the possible whistlers if they are not intense enough. 

\begin{figure}
\includegraphics[width=9.0cm]{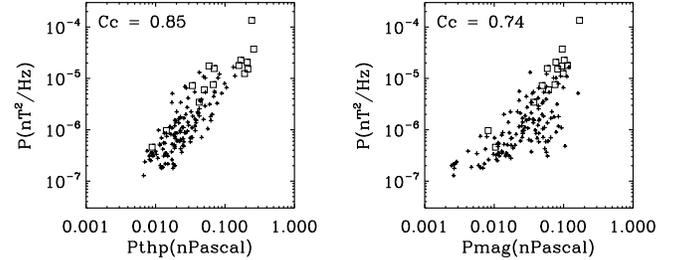}
\caption{Properties of the background turbulence. The power spectral density $P(f)$ (at $18$~Hz) in the 149 intervals without whistler waves (on Cluster~4) versus  the proton thermal pressure  $P_{thp}$ (left panel), and versus the magnetic pressure 
$P_{mag}$ (right panel). The squares represent intervals downstream of interplanetary shocks. $Cc$ is the correlation coefficient.}
\label{fig7}
\end{figure}


The second reason is that the turbulence level is correlated to $P_{thp}$ \citep{smith06b,alexandrova13a}. This is illustrated in the left panel of Figure~\ref{fig7}  where the turbulence intensity $P(f)$ at 18~Hz is drawn versus  $P_{thp}$ for 149 intervals without whistlers on Cluster~4: the correlation coefficient is $0.85$. The intensity of this turbulence can thus mask the possible whistlers when $P_{thp}$  is large, for instance in compression regions or downstream of shocks.
 Note that there are more than 40 points with $P_{thp} > 0.04$~nPa in Figure~\ref{fig:visib} (left panel): this is statistically significant for our data sample. Therefore, $P_{thp} = 0.04$~nPa can be considered as an upper limit of the ion thermal pressure for whistler observations.

What is the role played by the magnetic field strength $B_0$ or the magnetic pressure $P_{mag} = B_0^2/2 \mu_{0}$ on the visibility of the whistlers? The right panel of Figure~\ref{fig:visib}  shows that whistlers are observed for a range of $P_{mag}$ ten times larger than the range of $P_{thp}$ (left panel). Figure~\ref{fig7} (right panel) shows the dependence between the magnetic turbulence intensity $P(f)$ at 18~Hz and $P_{mag}$, with the correlation coefficient $Cc = 0.74$, which is slightly weaker  than the correlation between $P(f)$ and $P_{thp}$ (left panel). Thus, even if $P_{mag}$ probably plays a role in the turbulence intensity, we consider that, in our sample, the proton thermal pressure $P_{thp}$ 
is the best index of the turbulence intensity in the spacecraft frame, {\it i.e.} a better measure of a possible 
"occultation" of the whistlers by the usual permanent turbulence. 

Five IP shocks were observed on Cluster, in the free solar wind, from February 2001 to May 2005: 
one reverse shock (2003/02/27 around 14:23~UT) and four forward shocks (2001/02/20 around 02:00~UT, 2003/02/17 around 22:20~UT, 2004/01/22 around 01:35~UT, 2004/01/26 around 19:17~UT). Whistlers were only found from 01:55 to 02:00~UT on the 2001/02/20, around 5~Hz, upstream of a weak shock. Thus, the proximity of an IP shock does not favor the presence of whistlers, in our sample. Indeed, the usual solar wind turbulence is more intense downstream of IP shocks, and can  thus mask the whistlers (see black squares in  Figures~\ref{fig:visib} and~\ref{fig7}).

We conclude that the detection of whistler waves is easier when the intensity $P(f)$ of the usual turbulence is low, and 
when the solar wind speed is low. A low level of turbulence is thus a necessary condition for the observation of whistlers in our sample, but it is not a sufficient condition: we see in the left panel of Figure~\ref{fig:visib} that whistlers are not always observed, even  for $V_{sw} \leq$ 300~km/s and $P_{thp} \leq$ 0.02~nPa. 

We have also checked that the presence of the whistlers does not depend on the proton temperature anisotropy or the parallel proton beta $\beta_{p\parallel}=nkT_{p\|}/B^2/2\mu_0$ (not shown). It depends on the proton temperature $T_p$: a low $T_p$ favors the visibility of whistlers. This is related to the results of Figure~\ref{fig:visib} (left panel) because it is well known that there is a strong correlation between  $T_p$ and $V_{sw}$, as well as an evident correlation between $T_p$ and $P_{thp}$.

We shall now look for conditions, other than a low solar wind speed and a low proton thermal pressure, which allow the observation of whistler waves.

\section{Rotation of the large scale magnetic filed}

In this section and in the next section, we consider intervals much longer than 10 minutes during which whistlers appear 
suddenly while Cluster remains in the free solar wind: what are the solar wind properties which control this whistler appearance?

\begin{figure}
\includegraphics[width=9.0cm]{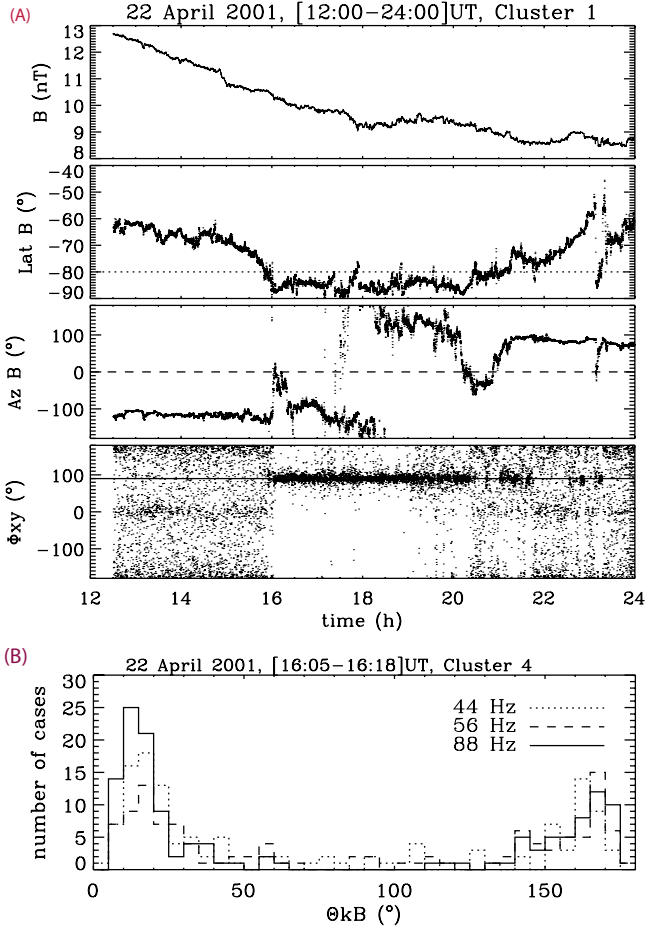}
\caption{(A) Upper panels: strength of the average ${\bf B_0}$ field, latitude and azimuth of the direction of ${\bf B_0}$. 
Lower panel: the phase difference between the $B_x$ and $B_y$ fluctuations at 44 Hz, in the whistler spectral peak. When $\Phi_{xy}$ is close to $90^{\circ}$ (horizontal solid line), the fluctuations are right-handed coherent whistler waves. 
(B) At three frequencies, histograms of the angle $\Theta_{kB}$ between ${\bf B_0}$ and 
the wave vector ${\bf k}$ of the whistler waves. }
\label{fig8}
\end{figure}


Let us consider an interval of 12 hours on day 2001/04/22, Figure~\ref{fig8}(A). The upper panels show the strength, the latitude and the azimuth of the ${\bf B_0}$ field. The lower panel shows 
the phase difference $\Phi_{xy}$ between the fluctuations measured in two orthogonal directions perpendicular to ${\bf B_0}$, at 44~Hz. 
 A more or less constant phase difference $\Phi_{xy} \simeq 90 ^{\circ}$ implies the presence of coherent whistlers  at this frequency, as we have seen in section~5. We note that the whistlers appear suddenly around 16:00~UT, when the azimuth of ${\bf B_0}$ jumps from $-120^{\circ}$ to $0^{\circ}$ (third panel). Then, the whistlers are permanent or intermittent until about 22:00~UT. Whistlers are mainly observed when the latitude of ${\bf B_0}$ is below $-80^{\circ}$ (horizontal dotted line in the second panel), {\it i.e.} when ${\bf B_0}$ is strongly southward. This southward latitude, as well as the smooth decrease of  $B_0$ (upper panel), indicate that Cluster is in a flux rope (Justin Kasper, private communication, 2014), 
after the crossing of the center of the rope which occurred earlier, around 08:00~UT.

For the time interval of Figure~\ref{fig8}(A), we have good measurements of the electric field fluctuations with Cluster/STAFF-SA at $f\ge 44$~Hz; therefore we can determine the sense of  the wave vector ${\bf k}$ without the $180^{\circ}$  ambiguity (as explained in section~2).  The histograms of  Figure~\ref{fig8}(B) show that  $\Theta_{kB}$ is observed to be 
around $10^{\circ}$ as well as around $170^{\circ}$, at 44~Hz (dotted line), 56~Hz (dashed line) and 88~Hz (solid line), 
during more than 10 minutes on the day of Figure~\ref{fig8}(A), in the flux rope. As noted in section 5, taking into account the solid angle of the gyrotropic wave vectors, we conclude that the wave vectors of the most intense 
whistlers can be parallel or antiparallel  to ${\bf B_0}$. As the GSE $B_X$ component is slightly positive during this interval, the waves with $\Theta_{kB} \approx 0^{\circ}$ propagate sunward, while those with $\Theta_{kB} \approx 180^{\circ}$ propagate antisunward. However, as the angle $\Theta_{BV}$ between ${\bf B_0}$ and the solar wind velocity is close to $90^{\circ}$, the Doppler shift of the whistlers is small, so that the sunward and the antisunward whistlers are seen at nearly the same frequency. 

The observation of waves in two opposite directions in a flux rope would be consistent with observations of 
bi-directional electron distribution functions. However, there are no electron data in this  time interval, which does not belong to our sample of  21 whistler intervals. 

According to  \citet{lin98} or to \citet{breneman10}, whistlers are observed near Stream Interaction Regions (SIR), or near  a crossing of the heliospheric current sheet (HCS). These regions are close to magnetic sector boundaries, through which the solar wind magnetic field polarity is reversed, so that the azimuth of ${\bf B_0}$ changes strongly. However, a strong (about $180^{\circ}$) change of the azimuth of ${\bf B_0}$ is only observed for about half of our intervals with whistlers. In the other intervals, Cluster did not cross the HCS; but it could have been close to it. Anyway, when the whistlers appear, there is always a change of the magnetic field direction, a change which can be small. 

The observed whistlers waves could be waves generated in a free solar wind region where a magnetic field reconnection 
occurs, propagating along ${\bf B_0}$, and reaching Cluster when it is magnetically connected to this reconnection region. We have found no way to test this hypothesis; but the facts that whistlers are related to SIRs 
and to the HCS, {\it i.e.} close to magnetic sector boundaries, and that they appear when the direction of ${\bf B_0}$ 
changes, support this hypothesis. Note that whistler waves have been observed on Cluster before and during the crossing of a magnetic reconnection region, in the Earth's magnetotail, see \citet{wei07} and references therein; see also the simulations of \citet{goldman14}.

\section{Electron distribution functions and whistler instabilities}

\begin{figure}
\includegraphics[width=9.6cm]{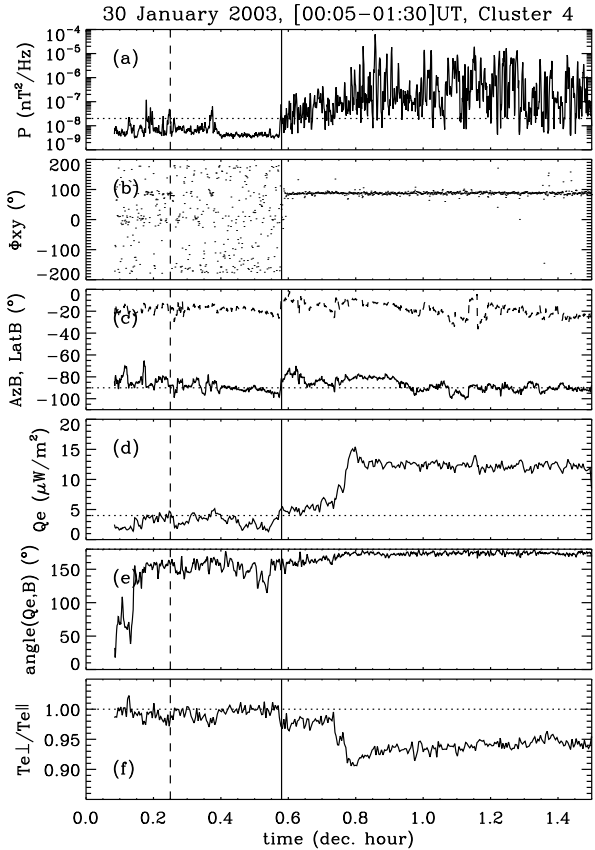}
\caption{During $1.5$~hour, at 88~Hz: (a) the power spectral density, (b) the phase difference  $\Phi_{xy}$.  (c) The latitude (dashed line) and the azimuth (solid line) of the direction of the magnetic field ${\bf B_0}$. (d) The heat flux $Q_e$, (e) the angle between ${\bf Q_e}$ and ${\bf B_0}$. (f) The anisotropy of the total electron temperature.  For the vertical lines, dashed and solid, see text.}
\label{fig10}
\end{figure}


The sources of whistler waves can be different instabilities, related to the electron distribution function: the electron firehose instability, when $T_{e\parallel}$ is larger than $T_{e\perp}$, the anisotropy instability when  $T_{e\perp}$ is larger than $T_{e\parallel}$, and the whistler  heat flux instability when the heat flux $Q_e$ is larger  than a critical value \citep{gary93book,gary99}.  Note that \citet{lin98} observe that the magnetic wave power of the whistlers around 14~Hz tends to increase when the electron heat flux increases.

In Figure~\ref{fig10}, we show an example of the presence of whistlers, in relation to the values of $T_{e\perp}/T_{e\parallel}$ and $Q_e$. On $2003/01/30$, from 00:00 to 01:30~UT, whistlers appear around 0.6~dec.hour (00:36~UT) (vertical solid line in Figure~\ref{fig10}) between about 35~Hz and 140~Hz: this is shown by the power spectral density $P(f)$ and by the phase difference $\Phi_{xy} \simeq  90^{\circ}$ at $88$~Hz (Figure~\ref{fig10}a and b). The interval 01:00$-$01:10~UT belongs to our sample of intervals with long-lived whistlers, and its spectrum is shown in Figure~\ref{fig2}a, the solid line peaking around 100~Hz.  

The whistler appearance is not due to a decrease of  $P_{thp}$ or of  $V_{sw}$ (see section~6): these two quantities (not shown) remain nearly constant. It is related to a small change of the direction of ${\bf B_0}$, (Figure~\ref{fig10}c), where the azimuth Az$B$ is shown by a solid line and the latitude Lat$B$ by a dashed line. The panel (d) gives the modulus of the total electron heat flux vector $Q_e$, in $\mu W/m^2$.  The panel (e) gives the angle $(Q_e,B)$ between the heat flux and ${\bf B_0}$. As discussed by \citet{salem01,salem03}, the vector $Q_e$ has to be parallel or anti-parallel to ${\bf B_0}$; but when the heat flux is small, its direction and intensity are poorly determined, owing to the spacecraft potential.  This happens in Figure~\ref{fig10}(d) and (e) between 0 and 0.6~dec.hour, when $(Q_e,B)$ is around $150-160^{\circ}$, and when $Q_e \leq$~4~$\mu W/m^2$. After 0.6~dec.hour, $Q_e$ is larger than 4~$\mu W/m^2$, the angle $(Q_e,B)$ reaches $160-170^{\circ}$, and the whistlers appear. This example shows that the  heat flux can be the source of the whistler instability. 

Another source for the whistlers could be the anisotropy of the electron temperature, $T_{e\perp}/T_{e\parallel} \geq 1$. But Figure~\ref{fig10}(f) shows that $T_{e\perp}/T_{e\parallel}$ is generally smaller than 1, and decreases slightly when the whistlers appear.

A heat flux instability is thus the probable source of the whistlers in the considered interval: whistlers appear around 0.6~dec.hour when $Q_e$ is larger than 4~$\mu W/m^2$; they are more intense after 0.75~dec.hour when $Q_e$ is larger than 12~$\mu W/m^2$ and the angle $(Q_e,B)$ larger than $170^{\circ}$. Even before 0.6~dec.hour, we note that small spikes of power spectral density (Figure~\ref{fig10}a) are observed around 0.25~dec.hour (see the vertical dashed line), with the whistler polarization (Figure~\ref{fig10}b), whenever the heat flux reaches 4~$\mu W/m^2$ (Figure~\ref{fig10}d).

\begin{figure}
\includegraphics[width=8.0cm]{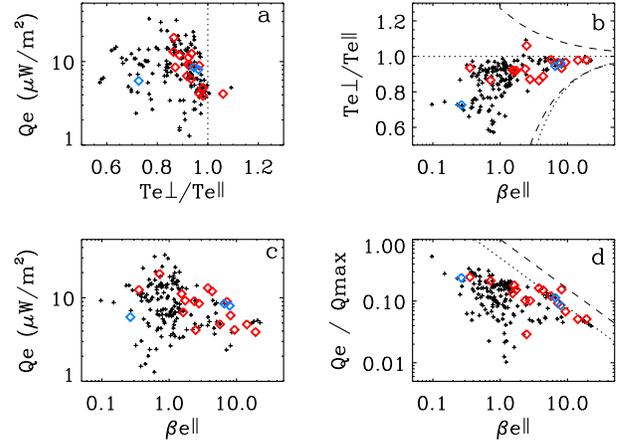}
\caption{Black crosses: electron properties for 149 events without whistler waves. The red (blue) diamonds are the 16  (3) intervals with intense (weak) whistler waves. (a) $Q_e$ versus $T_{e\perp}/T_{e\parallel}$. (b) $T_{e\perp}/T_{e\parallel}$ versus $\beta_{e\parallel}$; the upper dashed line gives an anisotropy threshold for the whistler anisotropy instability; the lower (dashed and dotted) lines give the thresholds for firehose instabilities (see text). (c) $Q_e$ versus $\beta_{e\parallel}$. (d) $Q_e/ 1.5 N_e k_B T_e V_{th,e}$ versus $\beta_{e\parallel}$; the dashed line  $1 / \beta_{e\parallel}^{0.8}$ \citep{gary99} is the threshold condition  $Q_e/ Q_{max}$ corresponding to a growth rate $\gamma = 0.01\omega_{ci}$ for the whistler heat flux instability; the dotted line $0.5 / \beta_{e\parallel}^{0.8}$ gives the upper bound of most of our data set, parallel to the threshold condition.}
\label{fig11}
\end{figure}


Among our sample of  21 intervals with well-established whistlers, 19 intervals have good enough measurements of the electron properties. We compare the electron heat flux and the electron temperature anisotropy of these 19 intervals to the 149 intervals without whistlers. We see in Figure~\ref{fig11}a  that the intense whistlers (red diamonds) are observed when 
$Q_e$ is  larger than 3 $\mu W/m^2$ and $T_{e\perp}/T_{e\parallel} \geq 0.8$.
There is only one case with $T_{e\perp}/T_{e\parallel}$ larger than 1, corresponding to whistlers.

As $T_{e\perp}/T_{e\parallel}$ is generally smaller than 1, can the firehose modes be unstable? Figure~\ref{fig11}b displays $T_{e\perp}/T_{e\parallel}$ versus $\beta_{e\parallel}$ = $2\mu_0 N_p k_BT_{e\parallel}/B_0^2$. The dotted line gives the anisotropy threshold, with a growth rate $\gamma=0.01\omega_{ci}$, for the electromagnetic non-resonant fire-hose instability: $T_{e\perp}/T_{e\parallel} = 1 - 1.70/\beta_{e\parallel}^{0.99}$ \citep{gary-nishimura03}. The unstable waves have a parallel propagation, but with a left-handed polarization, which is not observed. The lower dashed line gives the anisotropy threshold for the electromagnetic resonant  fire-hose instability: $T_{e\perp}/T_{e\parallel} = 1 - 1.23/\beta_{e\parallel}^{0.88}$ \citep{gary-nishimura03}. But the unstable waves are oblique, with a frequency equal to 0 in the plasma frame. Thus, even if these firehose instabilities constrain the electron temperature anisotropy \citep{camporeale08,hellinger14}, they cannot be the source of the observed  whistler waves. 

The upper dashed curve in Figure~\ref{fig11}b is the threshold for the whistler anisotropy instability 
$T_{e\perp}/T_{e\parallel} = 1 + 0.27/\beta_{e\parallel}^{0.57}$, still for $\gamma= 0.01\omega_{ci}$ \citep{gary-wang96}. The polarization and propagation properties of waves generated by this instability would be consistent with our observations but the observed anisotropy is usually too low.

We shall now consider the heat flux versus $\beta_{e\parallel}$ in our sample (we recall that we have not used the core and halo electron properties because the total electron temperature and the total heat flux are the only available data sets). Figure~\ref{fig11}c displays $Q_e$ versus $\beta_{e\parallel}$: there is an upper limit for  $Q_e$ which decreases  when $\beta_{e\parallel}$ increases. We draw the normalized heat flux $Q_e/Q_{max}$ versus $\beta_{e\parallel}$, where $Q_{max} = (3/2) N_e k_B T_e V_{th,e\parallel}$ is the free streaming heat flux, and $V_{th,e\parallel}= \sqrt{k_B T_{e\parallel}/m_e}$ the parallel electron thermal speed: Figure~\ref{fig11}d shows that $Q_e/Q_{max}$ is smaller than 0.3, a limit value frequently observed 
\citep{salem03,bale13}. More important, Figure~\ref{fig11}d shows that, when $\beta_{e\parallel}\geq 3$, a large part of the whistler events are close to an upper limit $0.5 / \beta_{e\parallel}^{0.8}$ (dotted line). This limit is parallel to the limit $1 / \beta_{e\parallel}^{0.8}$ (dashed line) given by \citet{gary99} for the upper bound of the normalized heat flux in the presence of a whistler instability with a growth rate $\gamma=0.01\omega_{ci}$. 

 The theoretical instability thresholds shown in Figure~\ref{fig11}b and Figure~\ref{fig11}d are based on simplified velocity distribution functions: an anisotropic Maxwellian core for the temperature anisotropy instability, and a core/halo model with a relative drift for the heat flux instability. As the solar wind electron distribution functions are more complex, the data are not expected to be constrained exactly by these theoretical thresholds.  Moreover, a different growth rate of the instability will shift the indicated theoretical threshold as well.  So, the complex non-thermal properties of the electron distributions and a weaker growth rate  could explain why the observed threshold is two times weaker than the theoretical prediction in Figure~\ref{fig11}d. 
Anyway, the upper bound of $Q_e/Q_{max}$ can be considered as related to the threshold of the heat flux whistler instability. Our modest sample of whistler intervals thus indicates that the whistler heat flux instability can play a role in the heat flux regulation: whistlers are indeed observed near the instability threshold, at least when $\beta_{e\parallel}$ is larger than~3.  Note that enhanced turbulent magnetic field fluctuations (around 0.3~Hz) along instability thresholds have been found by  \citet{bale09} for proton instabilities, see also \citet{wicks13b}.

Another instability than the heat flux instability can play a role in the generation of the whistlers. In
Figure~\ref{fig11}(d), we see that there is an interval with whistlers (red diamond) for $\beta_{e\parallel}=2.5$ 
and a weak heat flux $Q_e/Q_{max}=0.03$. This point corresponds to the red diamond with  $T_{e\perp}/T_{e\parallel}$ 
greater than~1  in Figures~\ref{fig11}(a) and (b): the whistler anisotropy instability could have played a role in this case.  

The frequencies and wavenumbers of the observed whistlers (see section~4) are consistent with those of the whistler 
heat flux instability (see \citet{gary93book}, Figure~8.8) as well as with those of the whistler anisotropy instability, see \citet{gary93book}, Figure~7.7.

However, a more precise description of the electron distribution functions, separating the core and halo temperature 
anisotropies, and the core and halo heat flux, would be necessary to study the growth rates of the 
considered whistler instabilities.
For instance, \citet{vinas10}  find that an electron strahl with a temperature anisotropy 
$T_{e\perp}/T_{e\parallel} \geq 2 $, observed on Cluster, could excite whistler waves above the 
lower hybrid frequency. They indeed find waves in an interval of (mainly) free solar wind with an anisotropic strahl, 
but they have not checked whether these waves were whistlers. Analysing the FGM CAA data, we find that these waves 
(not shown) are between $f_{ci}$ and $f_{lh}$, have a mainly linear polarization, and thus cannot be the whistler 
waves considered here.

\section{Role of the electron collisions}

The solar wind electron properties, the temperature anisotropy and the heat flux, are partly related to the Coulomb collisions between electrons \citep{salem03}; see as well the simulations of \citet{landi12,landi14}. The electron collisions thus probably play a role in  the generation and the visibility of the whistlers.  Following \citet{salem03}, we calculate the electron mean free path $L_{fp}$ for thermal electrons 
\begin{equation}\label{eq:Lfp}
L_{fp}  =  V_{th,e} / \nu_{ee}
\end{equation} 
\noindent where $V_{th,e}$ = $\sqrt{2 k_B T_e/m_e}$ is the electron thermal speed, and where $ \nu_{ee}$ in $s^{-1}$ 
is the basic electron collisional frequency for transport phenomena  
\begin{equation}
 \nu_{ee} \simeq 2.9 \times 10^{-6} N_e T_e^{-3/2} {\rm ln} \Lambda
\end{equation} 
\noindent with the Coulomb logarithm  ln$\Lambda \simeq 25.5$. 

We also consider the electron collisional age $A_e$ which relies on the $e-e$ thermal collisions which produce a transverse diffusion.  The corresponding collision frequency  in $s^{-1}$ is 
\begin{equation} 
\nu_{e\perp} \simeq 7.7 \times 10^{-6} N_e T_e^{-3/2} {\rm ln} \Lambda.
\end{equation}
\noindent $A_e$ is  the number of collisions suffered by a thermal electron between 0.5 and 1 AU \citep{salem03,stverak08}
\begin{equation}\label{eq:Ae}
 A_e \simeq 5.8 \times 10^4 N_e T_e^{-3/2} / V_{sw},  
\end{equation} 
\noindent where $V_{sw}$ is in km/s. In equations (\ref{eq:Lfp}) to (\ref{eq:Ae}), $N_e$ is in $cm^{-3}$ and $T_e$ in eV. 

\begin{figure}
\includegraphics[width=9.0cm]{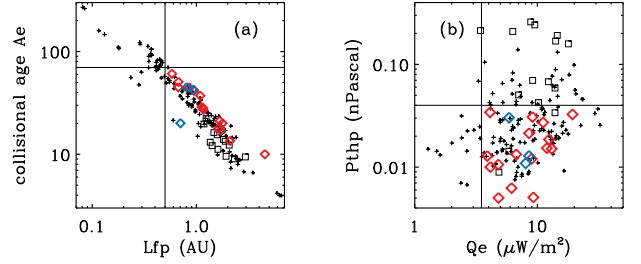}
\caption{For 168 intervals of 10 minutes. The red (blue) diamonds are the 16 (3) intervals with intense (weak) 
whistler waves. Black symbols indicate the intervals without whistlers: squares for the intervals downstream of interplanetary shocks, crosses for the other intervals. (a) Scatter plot of the electron collisional age $A_e$ and the mean free path $L_{fp}$: a necessary condition for the presence of whistler waves is $L_{fp} > 0.5$ AU. (b) Scatter plot of the proton thermal pressure $P_{thp}$ and the heat flux $Q_e$: two necessary conditions for the visibility or presence of whistler waves are $P_{thp} < 0.04 nPa$ and $Q_e > 3.5 \mu W/m^2$.}
\label{fig12}
\end{figure}


In Figure~\ref{fig12}a, we show $A_e$ as a function of $L_{fp}$ for intervals without whistlers (black crosses and squares) and for intervals with whistlers (red and blue diamonds). Whistlers are found for $L_{fp}$ larger than 0.5~AU (vertical solid line), and for a collisional age smaller than about 70 (horizontal solid line). Thus, whistlers are generated and visible in solar wind intervals with a few collisions. A large $L_{fp}$ is a necessary condition for the presence of whistlers, but it is not sufficient: whistler waves are present in only 15\% of the intervals with $L_{fp}$ larger than 0.5~UA.

\section{Discussion on possible other whistler sources}

We have found that the presence of whistlers was sometimes related to the local properties of the electron distribution 
function (section~8). Can some whistlers be related to other waves in the solar wind, or due to a non-local source?

Whistler waves in the solar wind have been observed in relation to Langmuir waves, giving Type III solar radio bursts \citep{kellogg92}, and in magnetic holes \citep{lin95,stone95}. Nevertheless, the WHISPER experiment shows that there are no Langmuir waves in our sample  (nor Type III bursts) when the whistlers are observed. Can the whistlers  be related to Langmuir waves themselves generated far from Cluster? Even if the whistler mode waves are able to propagate freely, far from the Langmuir waves, it is difficult to imagine that whistlers could be observed during 10 minutes, without any Langmuir wave (nor Type III radio signal) on Cluster. Thus, in our sample, the whistler waves are not related to Langmuir waves.

Can the whistler waves observed in the free solar wind be generated by non-local sources? It is well known that whistlers are sometimes present in the foreshock of the Earth's bow shock \citep{zhang98}. Can some of the free solar wind whistlers be foreshock whistlers, escaping from the foreshock along ${\bf B_0}$ field lines which are not straight lines? This is possible when a rapid change in the direction of ${\bf B_0}$ makes a connected field line to appear as a disconnected field line \citep{podesta13-2}. But such a transient geometry would only produce short-lived whistlers, not whistlers lasting several minutes. Furthermore, whistlers observed in the free solar wind and whistlers observed in adjacent regions of the foreshock generally have different frequencies and different directions of propagation. For instance, the free solar wind whistlers shown in Figure~1 between 17:15 and 17:42~UT, are observed between 7 and 40 Hz with an angle $\Theta_{kB}$ smaller than 15$^{\circ}$. Between 17:00 and 17:15~UT, foreshock whistlers are observed at higher frequencies, between 20 and 70~Hz, with a lower degree of polarization and larger angles $\Theta_{kB} $, from 5$^{\circ}$ to 25$^{\circ}$. On two other days (2004/02/22 around 10:00~UT, 2004/04/18 around 12:00~UT) foreshock whistlers are observed at frequencies higher than those of whistlers in the  adjacent free solar wind. Conversely, on the day 2003/01/30 (Figure~\ref{fig10}) foreshock whistlers around 01:55~UT (not shown) are observed at the same frequency ($88$~Hz) as the free solar wind whistlers; but these foreshock whistlers are related to a local strong heat flux of electrons backstreaming from the bow shock. Thus, in all these cases, the free solar wind whistlers cannot be due to the propagation of foreshock whistlers. 

However, as noted in section~7, the solar wind whistlers could be whistlers generated in magnetic reconnection 
regions of the free solar wind, and propagating along the ${\bf B_0}$ field. 

Are there different kinds of whistlers in the free solar wind? The results of \citet{zhang98}  for the direction of propagation of the whistlers are based on a minimum variance analysis of the magnetic field fluctuations. Similarly, our results are based on a SVD analysis of the same fluctuations. According to these two studies, the solar wind whistlers are quasi-parallel, with an angle  $\Theta_{kB}$ smaller than $15^{\circ}$. Conversely \citet{lengyel-frey94}, using the $B/E$ ratio of magnetic to electric field amplitude, find highly oblique whistlers downstream of IP shocks. \citet{breneman10}, using the three electric field components, find intense whistlers with a large electrostatic component and a highly oblique propagation, within stream interaction regions and near some shocks. \citet{lin98} find that the magnetic wave power of the whistlers tends to increase when the electron heat flux increases. (They also find mainly electrostatic waves, in the whistler frequency range, in regions with a reduced heat flux intensity, when the solar wind speed is decreasing.  
But \citet{lin03} note that this last result is not valid because the electric field noise below 10~Hz is contaminated by the spin modulation of the electric field caused by the photoelectron cloud around the spacecraft). Anyway, it seems that the whistler properties are different according as they are deduced from the electric field or the magnetic field observations. Different kinds of whistlers could be present in the free solar wind, some of them, more electrostatic, being not visible in the magnetic  fluctuations  studied here.


\section{Conclusion}

We have considered five years of Cluster data and selected a sample of 10-minute intervals in the free solar wind. The STAFF experiment gives continuously the intensity and the polarization of the magnetic fluctuations between 1~Hz and the electron cyclotron frequency, {\it i.e.} in the {\it whistler frequency range}. In this range, only 10\% of the considered intervals show the presence of long-lived (more than 5 minutes) right-handed whistler mode waves, with a quasi-circular polarization and a propagation quasi-parallel to the average magnetic field, in a narrow frequency band. These whistler bands are superimposed on the spectrum of the permanent non-polarized solar wind turbulence. Thus, coherent quasi-parallel whistlers waves do not seem to be ubiquitous in the free solar wind. 

 The fact that 10\% of our data set show the presence of whistler waves does not mean that these waves are present 10\% of the time in the free solar wind. The visibility of the waves depends on the solar wind properties. Indeed, we find that whistlers are observed for $V_{sw}<500$~km/s and for a low proton thermal pressure, $P_{thp}<0.04$~nPa (section 6).  For high solar wind speed and thermal pressure, the non-polarized background turbulence is intense and may hide possible whistler waves.  The fact that $P_{thp}$ is large downstream of the five interplanetary shocks observed by Cluster, which implies high turbulence level, can explain why whistler waves are not visible downstream of these IP shocks in our sample (section~6).

We find, as well, that whistlers appear  when there is a change of the magnetic field direction, a change which can be small (section~7). The quasi-parallel whistlers could be whistlers generated in regions of the free solar wind where a magnetic field reconnection occurs, and propagating along the ${\bf B_0}$ field, far from their source.

Another important condition of appearance of quasi-parallel whistler waves  is the presence of an electron heat flux $Q_e$ larger than 3 to 4 $\mu W/m^2$ (section~8). In section 9 we show that  a low collision frequency is also a necessary condition for the presence of quasi-parallel whistlers in the free solar wind (Figure~\ref{fig12}a). 

Figure~\ref{fig12}b illustrates the role of the two main necessary conditions: it shows the ion thermal pressure as a function of the electron heat flux. Among our sample of 10-minutes intervals, 25\% have a large enough heat flux, but whistlers should not be detected because $P_{thp}$ is too large (upper right quadrant): here, the quasi-parallel whistlers could be unstable, could play their part in the heat flux regulation, but  would not be visible in the spacecraft frame because the solar wind  turbulence is intense. In the lower right quadrant, we see that the two necessary conditions are not sufficient: about 100 intervals do not show the presence of whistlers, in spite of a large  heat flux and a low $P_{thp}$.

The generation of whistlers in the solar wind can be due to local sources, the anisotropy of the electron temperature, or a heat flux instability (section~8). We do not find indications that the temperature anisotropy instability plays an important part in the whistler generation; however, the available electron data only give the total electron temperatures, without separation between core, halo and strahl. Thus, we have not been able to check whether  a core or halo temperature anisotropy instability can be ruled out. Conversely, the fact that whistlers are precisely observed along the heat flux threshold of the whistler heat flux instability, when the electron parallel factor $\beta_{e\parallel}$ is larger than 3 (Figure~\ref{fig11}d), could imply that the whistler heat flux instability is at work, and contributes to the regulation of the solar wind heat flux, at least for $\beta_{e\parallel} \ge$ 3,  in the slow wind, at 1 AU. A better description of the electron distribution functions, separating the heat flux and the temperatures of a core, a halo and a strahl, would be necessary to study the growth rates of the considered whistler instabilities.

Finally, our identification of the whistlers is based on the magnetic field spectral matrix of STAFF-SA, calculated over 4~s. The phase difference $\Phi_{xy}$ is a powerful tool to detect the presence of whistlers. But $\Phi_{xy}$ has statistical uncertainties (see the bottom panel of Figure~\ref{fig8}(A)): $\Phi_{xy}$ has to be close to $90^{\circ}$ for several consecutive points to ascertain that whistlers are present. If whistler bursts last less than about 20 or 30~seconds (as the example shown in Figure~\ref{fig:sc-waves}), they shall not be identified by our method, which is appropriate for long-lived polarized fluctuation. To study whistler waves of any lifetime, one should complete our analysis with the STAFF-SC waveforms measurements in burst mode. In this case, $\Phi_{xy}$ can be determined with a resolution of a few tenths of second, from about 1 to 100~Hz. (However, the same high resolution for the electron data would be necessary to make a relation between intermittent whistlers and the electron distributions).

There are several unanswered questions about solar wind magnetic and electric fluctuations in the {\it whistler range}, and the corresponding electron properties,  which should be addressed in the future.

\acknowledgements
 We thank Pierrette D\'ecr\'eau and Patrick Canu for the use of the WHISPER data. We are grateful to Patrick Robert for a polarization program of waveform data, STAFF-SC and FGM. The FGM data (PIs A. Balogh and E. Lucek) and the CIS data (PIs  H. R\`eme and I. Dandouras) come  from the Cluster Science Data System (ESA). The PEACE data (PI A. Fazakerley)  come from the Cluster Active Archive (ESA). The French contribution to the Cluster project has been supported by the European Space Agency (ESA) and by the Centre National d'Etudes Spatiales (CNES).  Ondrej  Santol\'ik acknowledges support from grants GACR205/10-2279 and LH12231.  We thank the referee for very useful remarks.

\bibliography{bibliography_2014}

\end{document}